\documentclass[12pt,preprint]{aastex}
\usepackage{graphics}
\input epsf
\begin{document}

\title{Chain Galaxies in the Tadpole ACS Field}

\author{Debra Meloy Elmegreen \affil{Vassar College,
Dept. of Physics \& Astronomy, Box 745, Poughkeepsie, NY 12604;
elmegreen@vassar.edu} }

\author{Bruce G. Elmegreen \affil{IBM Research Division, T.J. Watson
Research Center, P.O. Box 218, Yorktown Heights, NY 10598, USA,
bge@watson.ibm.com} }

\author{Clara M. Sheets \affil{Department of Physics, Haverford College,
Haverford, PA, e-mail:csheets@Haverford.edu}}

\begin{abstract}
Colors and magnitudes were determined for 69 chain galaxies, 58
other linear structures, 32 normal edge-on galaxies, and all of
their large star formation clumps in the HST ACS field of the
Tadpole galaxy. Redshifts of 0.5 to 2 are inferred from
comparisons with published color-evolution models. The linear
galaxies have no red nuclear bulges like the normal disk galaxies
in our field, but the star formation clumps in each have about the
same colors and magnitudes. Light profiles along the linear
galaxies tend to be flat, unlike the exponential profiles of
normal galaxies. Although the most extreme of the linear objects
look like beaded filaments, they are all probably edge-on disks
that will evolve to late Hubble type galaxies. The lack of an
exponential profile is either the result of a dust scale height
that is comparable to the stellar scale height, or an
intrinsically irregular structure. Examples of galaxies that could
be face-on versions of linear galaxies are shown. They have an
irregular clumpy structure with no central bulge and with clump
colors and magnitudes that are comparable to those in the linears.
Radiative transfer solutions to the magnitudes and surface
brightnesses of inclined dusty galaxies suggest that edge-on disks
should become more prominent near the detection limit for surface
brightness. The surface brightness distribution of the edge-on
galaxies in this field confirm this selection effect. The star
formation regions are much more massive than in modern galaxies,
averaging up to $\sim10^9$ M$_\odot$ for kpc scales.
\end{abstract}


\section{Introduction}

Recent high resolution deep optical imaging of regions such as the
Hubble Deep Field North (HDF-N; Williams et al. 1996) and South
(HDF-S; Volonteri, Saracco, \& Chincarini 2000) and the Hawaiian
Deep Field (Cowie, Hu, \& Songaila 1995) has enabled the study of
large samples of non-local galaxies. Such fields showed that
galaxies with $z>0.5$ can have non-standard morphologies (van den
Bergh et al. 2000). The most linear have been called chain
galaxies (Cowie et al. 1995; van den Bergh et al. 1996), while
head-tail galaxies have been called tadpoles. Such objects may be
transient peculiar objects (Taniguchi  \& Shioya 2001) or an early
stage of normal galaxy evolution (Cowie et al. 1995). Previous
studies measured typical apparent I-band magnitudes of $\sim24.5$
mag and went as faint as $I=27$ mag (Lilly, Cowie, \& Gardner
1991). Here we present photometric and structural measurements of
127 linear objects out to 27 mag using the Hubble Space Telescope
(HST) field of the Tadpole galaxy (UGC 10214).

\section{Data and Analysis}

Deep images of UGC 10214 were obtained with the Advanced Camera
for Surveys (ACS) on HST in April 2002 (Tran et al. 2003). The
Wide Field Channel (WFC) provides a field of view of 3.4 arcmin x
3.4 arcmin with 0.05 arcsec resolution and the combined images
measure 3.9 arcmin x 4.2 arcmin. The exposure times were 13600 s
in the F475W filter (g band), 8040 s in the F606W filter (broad V
band) and 8180 s in the F814W filter (I band). The images were
kindly made available in advance of publication in fully reduced
form by the ACS team, using the methods of Blakeslee et al. (2003)
and Tran et al. (2003) but with more up-to-date reference files
and a new damped sinc ('lanczos3') drizzling kernel for improved
resolution over what is available in the Early Release
Observations (ERO) on the public website. Figure 1 shows a 1
arcmin x 1 arcmin portion of the ACS field.

Linear objects in the field were grouped into four morphological
categories: chain, having linear and clumpy structures (seen in
Figure 1); double, having two big clumps; tadpole, having a head
and tail shape, and normal, for edge-on spiral galaxies with
distinct bulges and occasional midplane dust lanes and clumps.
Representative objects are shown in Figure 2, with V-band
grayscale images and contours. The contours are in linear
increments; the lowest contour level is 1 $\sigma$ of the pixel
noise, which corresponds to a surface brightness of 25.5 mag
arcsec$^{-2}$.

The axial ratios of the objects were determined by measuring
lengths and widths from contour plots at a contour level 2
$\sigma$, corresponding to a surface brightness of 24.75 mag
arcsec$^{-2}$.  The chain, double, tadpole and normal galaxy
objects have average axial ratios of 5.1$\pm$ 2.0, 4.0$\pm$1.5,
3.7$\pm$1.8, and 3.2$\pm$1.0, respectively.

Total integrated magnitudes were determined in all three filters
and converted to the AB magnitude system, in which a galaxy with a
flat spectrum has equal magnitudes in each passband. We used $m =
-2.5 \log({\rm counts}/{\rm exposure\; time}) + {\rm zeropoint}$,
with the following zeropoints (see Blakeslee et al. 2003): 26.045
for g band, 26.493 for V band, and 25.921 for I band. Magnitudes
were determined from counts within a box outlining each object,
based on contour plots where the lowest contour is 1 $\sigma$.
Boxes were selected to be 5 pixels wider than the outer contours
of the objects, based on an examination of the turnover point on a
plot of magnitude as a function of box size.

The chain galaxies range from I=22 to I=27 mag, with six fainter
than 26 mag. For comparison, the Cowie et al. survey of the
Hawaiian deep fields was complete to about I=26 mag, and all 25 of
their chain galaxies had magnitudes between I=22 and I=25. The
only objects brighter than I=22 mag in our sample are normal
edge-on spiral galaxies.

Radial profiles were made along the major axis of each object
using strips 7 pixels wide. Sample profiles for g, V, and I band
intensity and (V-I) color are shown in Figure 3. The objects that
look like normal edge-on galaxies have exponential profiles with
central bulges. Chain galaxies, doubles, and tadpoles of the same
apparent size and magnitude as normals have neither exponential
profiles nor prominent bulges. Their interclump profiles are
nearly flat, and their clumps have similar brightnesses and colors
with no bright red object that could be a bulge (see also Cowie et
al. 1995, Figure 22).

Vertical marker lines in Figure 3 connect the intensity peaks in
the top part of the figure with color features in the bottom part.
The central peak in the normal galaxy is $\sim0.5$ mag redder than
its surroundings, while most of the peaks in the chain, double,
and tadpole objects are $\sim0.2$ mag bluer than their
surroundings.  The normal galaxy also has a blue off-center clump.

In Figure 4 (top), color-color diagrams are shown for the
integrated light of all of the galaxies. The chain, double, and
tadpole galaxies fill the same blue region of the diagram. Some of
the normal galaxies are also blue, while others are red.
Morphologically, the normal galaxies all appear similar, with a
central bulge and an exponential disk. It is possible that the
redder normals are earlier-type galaxies and the blue ones are
later type. There could also be a blueing effect with distance
(e.g., Lilly et al. 1991), assuming the smaller and bluer galaxies
are more distant on average.  The crosses are not linear galaxies
like the rest but are clusters of clumps that could be face-on
versions of the linear galaxies, as discussed in Section
\ref{sect:model}.

Figure 4 (bottom) shows the distribution of color for all of the
clumps in 8 representative chain galaxies (open symbols) and 7
representative normal edge-on galaxies (filled symbols). Double,
tadpole, and all of the other chain galaxies in our study have
similar clump colors and are not plotted to avoid confusion. The
triangles in the figure are integrated magnitudes, the circles are
non-central clumps, and the squares are central clumps. The
central clumps of the normal edge-on galaxies are red with
(V-I)$>1$; they are probably normal bulges. The non-central clumps
in all of the galaxies and the central clumps in the chain
galaxies are all somewhat blue and are most likely regions of
recent or active star-formation.  Clump clusters are also shown as
cross and plus symbols.

The surface brightnesses of the galaxies were determined from
their isophotal magnitudes and their sizes based on contour plots.
For chains, doubles, and tadpoles, the average I-band surface
brightnesses are $23.43\pm0.83$, $23.28\pm0.66$, and
$23.55\pm0.54$ mag arcsec$^{-2}$, respectively, while for normal
edge-on galaxies the average is slightly brighter, $22.91\pm0.91$.
Figure 5 shows the distribution of surface brightness for all of
the galaxies.  The vertical line is the $3 \sigma$ detection
limit, taken to be 3 times the rms deviation of a Gaussian fit to
the pixel noise in a relatively blank $100\times100$ pixel field
of view.  The count of galaxies begins to drop as the surface
brightness limit is approached, suggesting there are many more
galaxies that could not be detected. However, the dominance of the
linear galaxies approaching this limit also suggests this peculiar
class might be only the brightest members in a distribution of
normal galaxies, selected preferentially here because they are
edge-on. A radiative transfer model in Section \ref{sect:model}
supports this possibility: face-on versions of the same galaxies
would be fainter and not so easily detected. The two crosses in
this figure are candidates for face-on versions, shown later in
Figure 9; they are indeed very low in average surface brightness.

The I-band apparent magnitude distributions of the whole galaxies
are shown in the top of Figure 6, and the I band magnitude
distributions of the individual clumps are shown in the bottom.
For the clumps, the number distributions peak at the same
magnitude for the chain, double, and tadpole galaxies and for the
non-central clumps in the normal galaxies, whereas the central
clumps (bulges) in the normal galaxies are $\sim3-4$ mag brighter
than the rest. This result again suggests that all of the clumps
in chain, double, and tadpole galaxies and the non-central clumps
in normal galaxies are similar. The clumps in the clump clusters
are similar too. The central clumps in chain galaxies are not
bulges because they have the same blue colors and the same faint
magnitudes as the other clumps in these galaxies. The whole
galaxies are brighter than the clumps by $\sim3$ magnitudes, and
the normal edge-on galaxies are brighter than the chain, double
and tadpole galaxies by $\sim2$ magnitudes, although there is a
faint tail of normal galaxies that extends to the realm of the
chains ($\sim25$th mag). The average magnitude of a chain, double
or tadpole galaxy is comparable to the average magnitude of a
normal galaxy bulge.

The fraction of galaxies of each morphological type is shown in
Figure 7 as a function of I-band magnitude. Normal galaxies
dominate the sample out to magnitude 24. Chain galaxies dominate
at fainter magnitudes and eventually account for about 40\% of the
sample. The fractions of each type are fairly constant beyond
magnitude 24.

Scaling the number of galaxies in our field to a square degree,
there are $1.5\times10^4$ chain galaxies per square degree, $9300$
doubles, $3200$ tadpoles, and $6800$ normal galaxies out to our
limit of 27th mag. For comparison, Lilly et al. (1991) count in
the HDF $2.5\times10^5$ galaxies per square degree out to 27th
mag, or $10^5$ galaxies per square degree out to 26th mag
(consistent with the B band galaxy counts in another field by
Metcalfe et al. 1995). Thus $\sim6$\% of galaxies out to 27th mag
should be chains. For our normal edge-on galaxies, we see the
equivalent of $6000$ galaxies per square degree out to 25th mag,
whereas Lilly et al. give $3\times10^4$ galaxies per square degree
of all inclinations and types to that magnitude. That means about
20\% of their sample should be normal galaxies seen edge-on.

Westera et al. (2002) modeled the z-dependence of the observed
color and magnitude of a hierarchically accreting protogalaxy. We
infer from their Figures 1 and 8 that all of our objects,
including the normal galaxies, have total stellar masses of
$\sim10^{10}$ M$_\odot$ and that the individual clumps have masses
less than or comparable to $\sim10^9$ M$_\odot$.  The angular
sizes of the chain galaxies are $\sim1$ arcsec, which corresponds
to $\sim5$ kpc if the redshift is $z\sim~1$. Thus, the chain,
double, and tadpole galaxies could all be normal late-type disk
protogalaxies, viewed edge-on and possibly interacting (for the
tadpoles), and with a few very massive star-formation clumps and
no old central bulge.

\section{Comparison with other Work}

The chain galaxies in our sample resemble some of the chain
galaxies previously reported in other fields, such as object
SSA22-16 in Cowie et al. (1995) and object HDF 3-531 in van den
Bergh et al. (1996). Koo et al. (1996) obtained spectra for 3
chain galaxies in their field studies; they had I magnitudes of
23.5 to 23.7 mag and redshifts from 0.8 to 1.2. The I magnitudes
of our chain galaxies range from 22 to 27, with the majority
fainter than 25, so our galaxies may be mostly beyond z=1.
Consistent with this result is the observation that the HST Medium
Deep Survey (Abraham et al. 1996) had very few chain galaxies at
I$<22$ mag. The evolutionary models by Cowie, et al. 
(1996) for B magnitude as a function of redshift also suggest that
our redshifts are between 1 and 2. Koo et al.'s plot of (V-I)
versus redshift as a function of galaxy type suggests that our
chain, tadpole, and double galaxies are in the color realm of
late-type spirals between z=1 and 2 (V-I = 0.3 for Sm galaxies at
z=1.5 to 2, or V-I = 0.9 for Sbc in that z range). This conclusion
is also consistent with the plot of absolute I mag vs. z for
different apparent magnitudes as a function of galaxy type by
Aguerri \& Trujillo (2002).  Thus, the integrated colors and
magnitudes of chain galaxies are consistent with expectations for
young normal galaxies at $z\sim1-2$.

The linear objects in our study do not have exponential disks and
are therefore not like nearby late-type edge-on spirals. For
example, the low surface brightness galaxy UGC7321 is an edge-on
galaxy with no apparent bulge, but it has an obvious exponential
disk (Pohlen et al. 2003). Some faint edge-on and face-on field
galaxies at $z=0.5$ to 1.5 also have exponential profiles (O'Neil,
Bothun, \& Impey 2000).

Our chain galaxies could be more irregular edge-on low surface
brightness (LSB) galaxies (Dalcanton \& Schectman 1996). The
surface brightness in the center of an LSB galaxy is $\sim23.4$
mag arcsec$^{-2}$ (McGaugh \& Bothun 1994), and midway out in the
disk it is 24.5 to 25.5 mag arcsec$^{-2}$. Our chain galaxies have
surface brightnesses midway out that are 23.5 to 24 mag
arcsec$^{-2}$, which is about the same as our normal galaxies and
$\sim1$ mag brighter than an LSB galaxy.  However, an edge-on LSB
galaxy should be brighter than a face-on LSB galaxy by more than a
magnitude per arcsec$^2$, as shown in section \ref{sect:model}
below, so the irregular-LSB explanation is possible.

A chain galaxy studied by Steidel et al. (1996), C4-06, has an
isophotal AB magnitude of 23.48 and an aspect ratio of 5:1, making
it similar to those studied here. The redshift of C4-06 is 2.8,
and it has very strong absorption lines that match those of a
redshifted spectrum of the starburst galaxy NGC 4214.
Unfortunately, the spectrum was too faint to detect rotation.

The structure of our chain galaxies resembles the beaded filaments
seen in N-body simulations of the early universe and of globular
cluster formation at z=4 and 7 (Kravtsov \& Gnedin 2003), as well
as some chains of galaxies imaged by Vorontsov-Velyaminov (1977;
e.g. VV144=Arp 151 at z=0.02) and Hickson (1993, see also
Mendes de Oliveira \&  Hickson 1994).
However, the mass and size of a globular cluster is small for our
clumps and the chains of galaxies in Vorontsov-Velyaminov are too
large for our systems.

Reshetnikov, Dettmar, \& Combes (2003) measured the length/width
ratios and radial and vertical profiles for a sample of 34 edge-on
galaxies in the Hubble Deep Field (HDF). They found that galaxies
at $z\sim1$ have average major/minor axial ratios of 3.3 and
concluded that the disks were about 1.5 times thicker than in
local galaxies. Half of their sample, including some face-on
galaxies, had non-exponential radial profiles as if they were
still in the process of disk formation. This conclusion is
consistent with galaxy evolution models, which show that disks may
not start to develop exponential radial profiles until $1<z<2$
(Westera et al. 2002). However, in most of the HDF edge-on
galaxies, a central bulge-like peak is prominent even if the disk
is not yet exponential. This differs from our chain galaxies,
which do not generally have central peaks. The Reshetnikov et al.
edge-on galaxies also differ in axial ratio from our chain
galaxies, which are about a factor of 2 thinner, although the
normal edge-on galaxies in our sample have the same axial ratio as
in Reshetnikov et al.. The sizes of galaxies in the two samples
are about the same: Reshetnikov et al. consider galaxies with
diameters larger than 1.3 arcsec, and this is the average size of
our chain galaxies.  Our doubles and tadpoles are slightly
smaller, averaging 1.1 arcsec, and our normal galaxies are
slightly larger, averaging 1.7 arcsec.

The disk galaxy evolution accretion models by Westera et al.
(2002) show a radial color gradient in (V-I) of about 0.2 mag from
center to edge at z=0, with a radial gradient of about 0.6 mag at
z=1.3. The (V-I) colors of the central red dip and disk for our
edge-on galaxy shown in Figure 1 are consistent with their model
for z=0.25. However, our chain galaxy in Figure 1 matches none of
the model color radial profiles, because no color gradient is
present. This is further evidence that our chain galaxies are not
exponential disk systems. Cowie et al. (1988) define a flat
spectrum population of galaxies with integrated colors
(B-I)$_{AB}<0.8$. The majority of our chain, double, and tadpole
objects have such a spectrum.

\section{Model}
\label{sect:model}

While chain galaxies have sizes, magnitudes, and colors comparable
to faint normal late-type galaxies in the same field, the apparent
dominance of the chains and other linear galaxies at faint
magnitudes and their lack of exponential profiles are surprising.
For random orientations, circular disk galaxies should have equal
numbers in equal intervals of axial ratio ($<1$). However, for
galaxies close to the surface-brightness limit of a survey, this
is not the case. Edge-on galaxies that are not too optically thick
have larger surface brightnesses than face-on galaxies of the same
type because of the longer path lengths through edge-on disks.
Edge-on galaxies have fainter total magnitudes because of their
smaller projected areas. Thus face-on versions of these linear
galaxies could be mostly below the surface-brightness detection
limit of our survey.

The exponential nature of disks is best seen for nearly face-on
galaxies where the line-of-sight extinction is small. Edge-on
galaxies can have flat intensity profiles if extinction is
important and we view only the near side of the disk. If the
linear galaxies are edge-on bulge-less disks, then they could be
either irregular with no intrinsic exponential light distribution,
or exponential with a dust scale height comparable to the stellar
scale height.

Edge-on galaxies have more extinction but their average surface
brightnesses can still be brighter than face-on galaxies because
the extinction path length is usually larger than the disk
thickness. Holmberg (1958) defined a corrected face-on surface
brightness as the total flux divided by the square of the
semimajor axis. If there is no internal extinction, then this is
independent of inclination. The actual surface brightness is the
flux divided by the projected area, rather than the deprojected
area. The corrected face-on surface brightness is brighter than
the actual average surface brightness by the extinction effect,
and it is fainter than the actual average surface brightness by
the ratio of axes.

Solanes, Giovanelli \& Haynes (1996) determined the variation of
corrected face-on surface brightness with axial ratio $b/a<1$
using the Holmberg definition $\Sigma_H=m_0+5\log a$ for semimajor
axis $a$ in arcsec. Here $m_0$ is the total magnitude corrected
for foreground Milky Way absorption. Solanes et al. obtained the
result $\Sigma_H=\Sigma_0+k\log\left(b/a\right)$ where the slope
of the relation is $k=-0.9,$ $-0.9,$ $-1.3,$ $-1.3,$ and $-1.4$
for Hubble types Sa, Sab, Sb, Sbc, and Sc, respectively. This
variation with $b/a$ is from extinction alone, indicating that
inclined galaxies are more heavily self-absorbed. The actual
average surface brightness is
$\Sigma_a=m_0+2.5\log\left(ab\right)$ for projected area $ab$.
Thus the actual $\Sigma_a$ is related to the Holmberg $\Sigma_H$
as $\Sigma_a=\Sigma_H+2.5\log\left(b/a\right)$.  The dependence of
actual average surface brightness on projected axial ratio is
$\Sigma_a=\Sigma_0+k^\prime\log\left(b/a\right)$ where
$k^\prime=k+2.5.$  The first term, $k<0$, is from internal
extinction and the second term, 2.5, is from the ratio of the
areas of the projected ellipse to the deprojected circle. For the
five Hubble types, $k^\prime= 1.6,$ $1.6,$ $1.2,$ $1.2,$ and
$1.1$, respectively. As $\left(b/a\right)$ decreases for more
highly inclined galaxies, the actual surface brightness in mag
arcsec$^{-2}$ decreases and surface brightness becomes brighter.
This makes galaxies close to the surface brightness limit of a
survey more easy to see if they are highly inclined.

Figure 8 shows a radiative transfer model of a disk with an
exponential profile in 4 scale lengths for both emissivity and
absorption. The top, middle, and bottom curves of each type have
perpendicular opacities to the center midplane equal to 0.33,
1.33, and 1.77 optical depths (1.33 gives an absorption rate of
1.5 mag/kpc at 2 scale lengths out in the disk if the scale length
is 3 kpc, and this is typical in V band for a modern galaxy). In
the bottom figure, the solid lines are the radial profiles along
the midplane for the disk viewed edge-on, the dashed lines are
along strips parallel to the midplane but displaced by one disk
thickness scale length (i.e., off-axis strips), and the dotted
lines are for face-on versions of these same disks. Brightness is
normalized by setting the volume emissivity to 1 in the center
midplane.  The natural log of the intensity is plotted; this is
approximately the same as a magnitude scale but with brighter
emission more positive in this case. The model assumes a vertical
profile of both emissivity and dust that is Gaussian with a $1/e$
half-thickness equal to 4\% of the disk scale length. Evidently,
the midplane and off-axis profiles are nearly flat, especially for
high extinctions, and the face-on profile is exponential with a
surface brightness that is lower by 2-3 mag.  Thus it is difficult
to tell if an edge-on galaxy has an exponential disk when the dust
thickness is comparable to the stellar thickness. If the dust
layer is thinner than the stellar layer, as in modern galaxies,
then the midplane alone may be dark but the exponential profile
will still appear along a major-axis strip that is displaced from
the midplane.

The ratio of the upper two opacities in these curves, 0.75, is the
same as the ratio of V to B extinction in the galaxy rest frame
for a normal extinction curve (where $A_V=3E(B-V)$). This
translates approximately to an extinction $A_I\sim3E(V-I)$ at the
expected redshift of these objects.  Thus the (V-I) color excess
from extinction ($=A_V-A_I$) for an edge-on disk of this type is
the difference between the two nearby curves of each type in
Figure 8, which is $\sim0.25$ mag.  The large range of observed
colors (V-I) in Figure 4 makes it difficult to tell if this small
amount of extinction is present in our sample.

The top of Figure 8 shows the average integrated surface
brightnesses of the same three galaxy models, now viewed at
inclination angles given by the log of the ratio of axes, $a/b>1$,
on the abscissa. The surface brightness becomes large at high axis
ratio because that projects the longest path length through the
stars, even with interstellar dust.  The average slope of this
relation in the figure, $k^\prime\sim1-1.5,$ is about the same as
the observed slope from Solanes et al. (1996) after conversion of
their Holmberg surface brightness to actual surface brightness
(see above). If the practical detection limit for the visual
identification of a linear structure is one magnitude of surface
brightness fainter than the edge-on value, as suggested by the
shift between the peak for the chain galaxies and the detection
limit in Figure 5, then Figure 8 gives us the expected range for
the ratio of axes of these galaxies.  This magnitude difference
corresponds to an $\ln({\rm average \;Intensity})$ that is 1 less
than the peak on the right of Figure 8.  In this case nearly
everything that can be seen will have an axis ratio greater than
3-4, the corresponding value on the abscissa. The inverse of this
ratio, $0.25-0.3$, would be the fraction of all comparable
galaxies that are detected. For this model, the chain, double, and
tadpole galaxies could be edge-on late-type or irregular galaxies
whose face-on counterparts are not obvious because they are below
the surface brightness limit of the survey.

To check whether there could be a faint population of galaxies
that are face-on versions of the chain, double, or tadpole
galaxies, we searched the HST field of view for small clusters of
emission clumps having about the same overall angular size and
clump count as a typical chain galaxy. Figure 9 shows two
examples.  The colors of these clump clusters and their individual
clumps are plotted in Figure 4 as crosses and plus signs,
respectively, the average surface brightnesses are plotted in
Figure 5, and the magnitudes of the clump clusters and individual
clumps are plotted in Figure 6.  Evidently, the integrated and
clump properties are similar to those of the chains, doubles and
tadpoles, but the average surface brightnesses of the clump
clusters are $\sim1$ mag arcsec$^{-2}$ fainter. The lowest
contours in Figure 9 are $1 \sigma$ for the left-hand clump
cluster, as in the contours of Figure 2, and $0.5\sigma$ for the
right-hand cluster.

A problem with the interpretation that chain galaxies are all
edge-on clumpy disks is that some of them in this survey, such as
the four shown in Figure 1, have average surface brightnesses that
are $\sim10\sigma$ of the image noise. These examples are not
close to the detection limit and yet no peripheral emission can be
seen that resembles the rest of an edge-on disk.  They would have
to be highly inclined disk galaxies if this model is correct.
Rotation curves or inclination statistics might eventually confirm
this.  Another problem is that the double and tadpole galaxies
would look this way from nearly any orientation.  They could
already be face-on and we would not know it if the underlying
galaxy is too faint to see.  Because the double and tadpole
galaxies have a peculiarity that cannot be transformed away by
projection effects, the chain galaxies might have an intrinsic
peculiarity too; i.e., they might really be chains even though
this structure is dynamically unstable.

The normal edge-on galaxies in our sample have bulges like
early-type galaxies and they have exponential profiles presumably
because their gaseous and stellar disks have relaxed and their gas
disks are thinner than their stellar disks. This relaxation makes
sense for early-type galaxies because their evolution is faster
than late-type galaxies.  If the chains, doubles and tadpoles are
disk galaxies too, then they either have no exponential profiles
or their exponential profiles are hidden by dust, as shown in
Figure 9. This would require their gaseous disks to be comparable
in thickness to their stellar disks. Such a large scale height for
dust in late-type galaxies is consistent with the very large
star-forming regions that are seen in these galaxies, because the
Jeans length in the ambient interstellar gas is usually comparable
to the scale height.

\section{Conclusions}

We measured the properties of 127 linear galaxies with chain,
double, and tadpole morphologies, as well as the properties of
other linear objects that are normal edge-on spirals, out to
I-band magnitude 27 in the HST ACS Tadpole field. We also measured
2 clump clusters that could be face-on versions of the chain
galaxies. All of the objects contain similar blue clumps that are
apparently massive star-forming regions, with masses up to $10^9$
M$_\odot$. The normal galaxies have centralized, red clumps that
are probably bulges, unlike the linear galaxies which have few red
clumps anywhere. The normal galaxies also have exponential
brightness profiles, unlike the linears which are flat.  The
chain, double and tadpole galaxies dominate all edge-on galaxies
in the field beyond an apparent magnitude of $\sim24$.

We considered that most chain galaxies could be edge-on, late-type
or low surface brightness galaxies that are close to the surface
brightness limit of our survey.  The late type explains the lack
of a bulge, while the surface brightness limit explains the
dominance of edge-on orientations (since edge-on galaxies have
higher surface brightnesses than face-on galaxies). We cannot tell
if the intrinsic stellar surface density profiles are exponential
with obscuring dust layers or if they are intrinsically
non-exponential. The clump clusters have non-exponential profiles
because their light distributions are dominated by the clumps.

If the chain and other linear galaxies are edge-on clumpy disks,
then a deeper survey should show more features in these galaxies,
making them look more like normal galaxies, while a new set of
linear galaxies should begin to dominate the field close to the
new surface brightness limit. On the other hand, if chain galaxies
look the same at higher sensitivity, then they are probably not
edge-on disks but filaments in a more exotic and transient phase
of galaxy formation. The total luminosity of a chain galaxy is
comparable to the luminosity of a normal galaxy bulge in our
sample, so if chain galaxies are unstable clumpy filaments (Cowie
et al. 1995) then they could collapse into bulge-like objects,
possibly providing seeds for normal galaxy formation.

Acknowledgments

We are very grateful to J. Blakeslee, H. Ford, and J. Mack of the
ACS team for providing up-to-date fully reduced combined images in
advance of public release. We thank the Keck Northeast Astronomy
Consortium for summer undergraduate support for C. Sheets. The
contribution by B.G.E. was partially funded by NSF Grant
AST-0205097.

\newpage
\begin{figure}
\caption{V-band image of a 1' x 1' field within
the Tadpole ACS field shows several objects classified here as
chain galaxies. (see 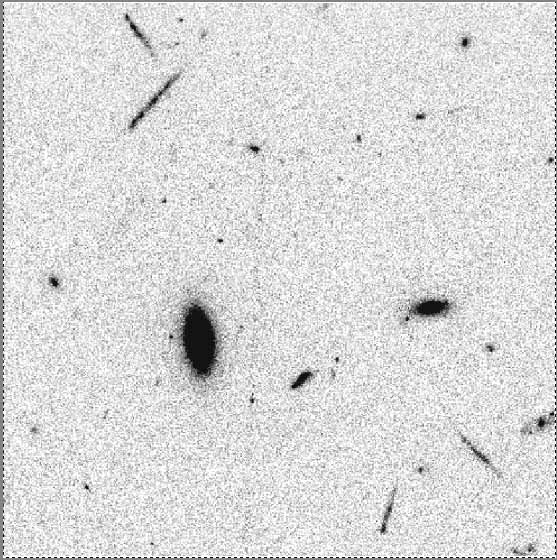)}\end{figure}

\begin{figure}
\caption{Morphologies are shown in V band for (a)
chain, (b) double, (c) tadpole, and (d) normal galaxy classes. The
angular scales are shown by the axis ticmarks, which are separated
by 10 pixels ($=0.5$ arcsec) for (a), (c), and (d), and 1 px
(=0.05 arcsec) for (b). (see 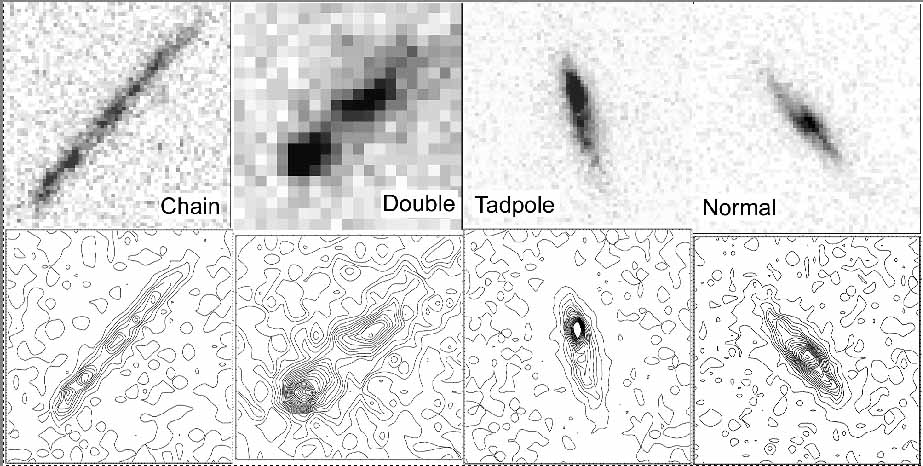)}\end{figure}

\begin{figure}
\plottwo{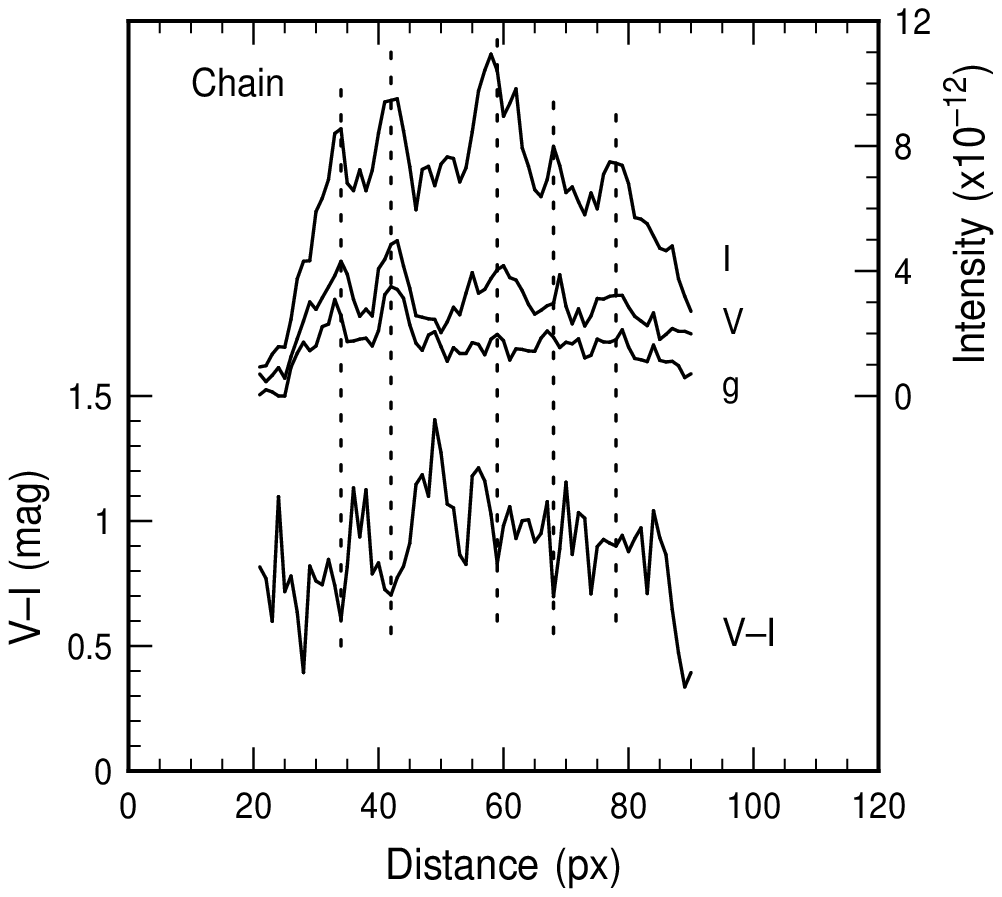}{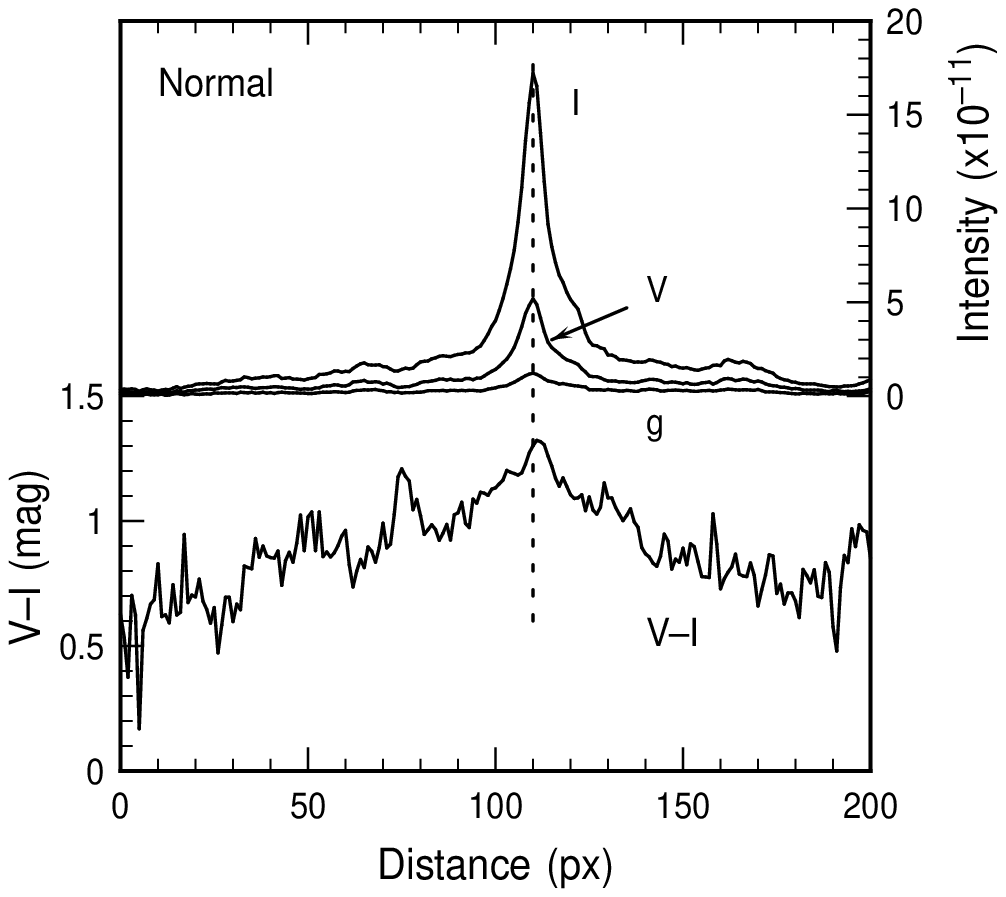} \caption{Radial profiles showing the
intensity (top) in each of the three bands as a function of
distance along the object (in px = 0.05 arcsec) and (V-I) mag
(bottom) for a chain galaxy (left) and normal galaxy (right).}
\end{figure}

\begin{figure}
\plotone{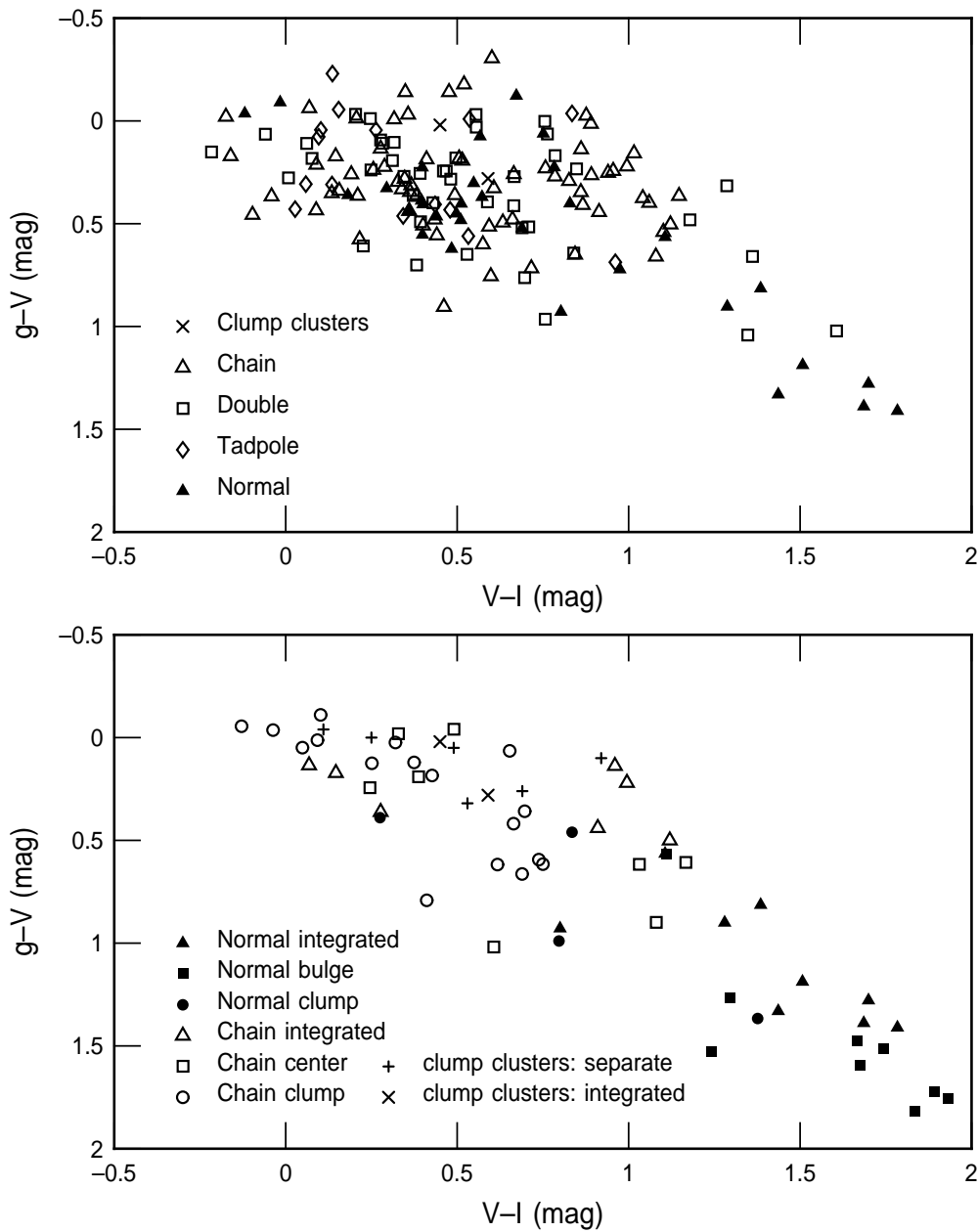} \caption{(top)Color-color diagram for whole
galaxies, with open symbols for the integrated chain, double, and
tadpole galaxies, and filled symbols for the integrated normal
galaxies.  The two crosses are the clump clusters in Fig. 9.
(bottom) Color-color diagram with filled symbols for a
representative sample of integrated normal galaxies, their central
peaks and off-center clumps, and open symbols for a sample of
integrated chain galaxies and their central and off-center clumps.
The crosses and plus signs are for the clump clusters in Fig. 9.}
\end{figure}

\begin{figure}
\plotone{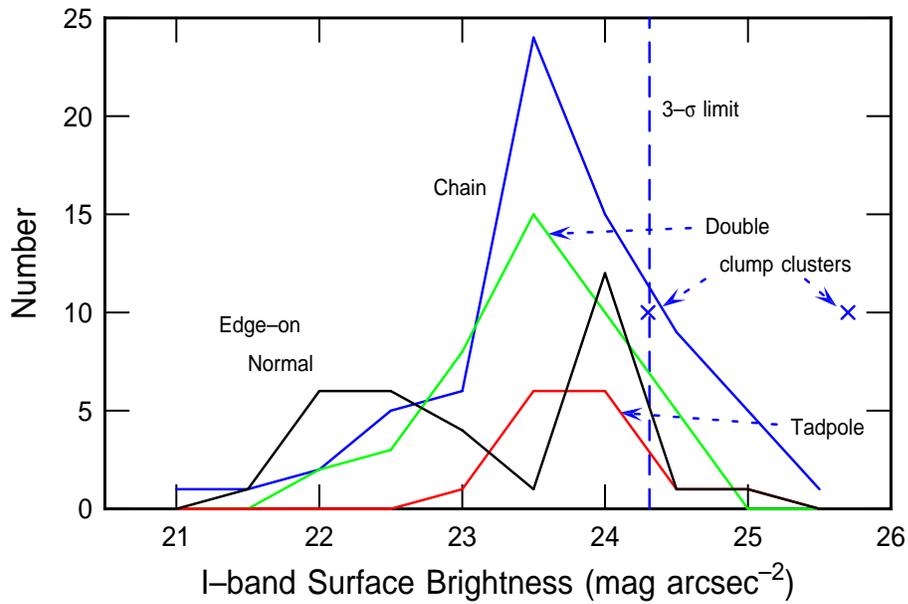} \caption{Distribution of surface brightness for
various morphologies. The vertical dashed line is the $3 \sigma$
detection limit. Chain galaxies dominate our sample at low surface
brightness. The galaxy count drops close to and beyond the $3
\sigma$ limit.  The two crosses indicate the average surface
brightnesses of the clump clusters shown in fig. 9.}\end{figure}

\begin{figure}
\plotone{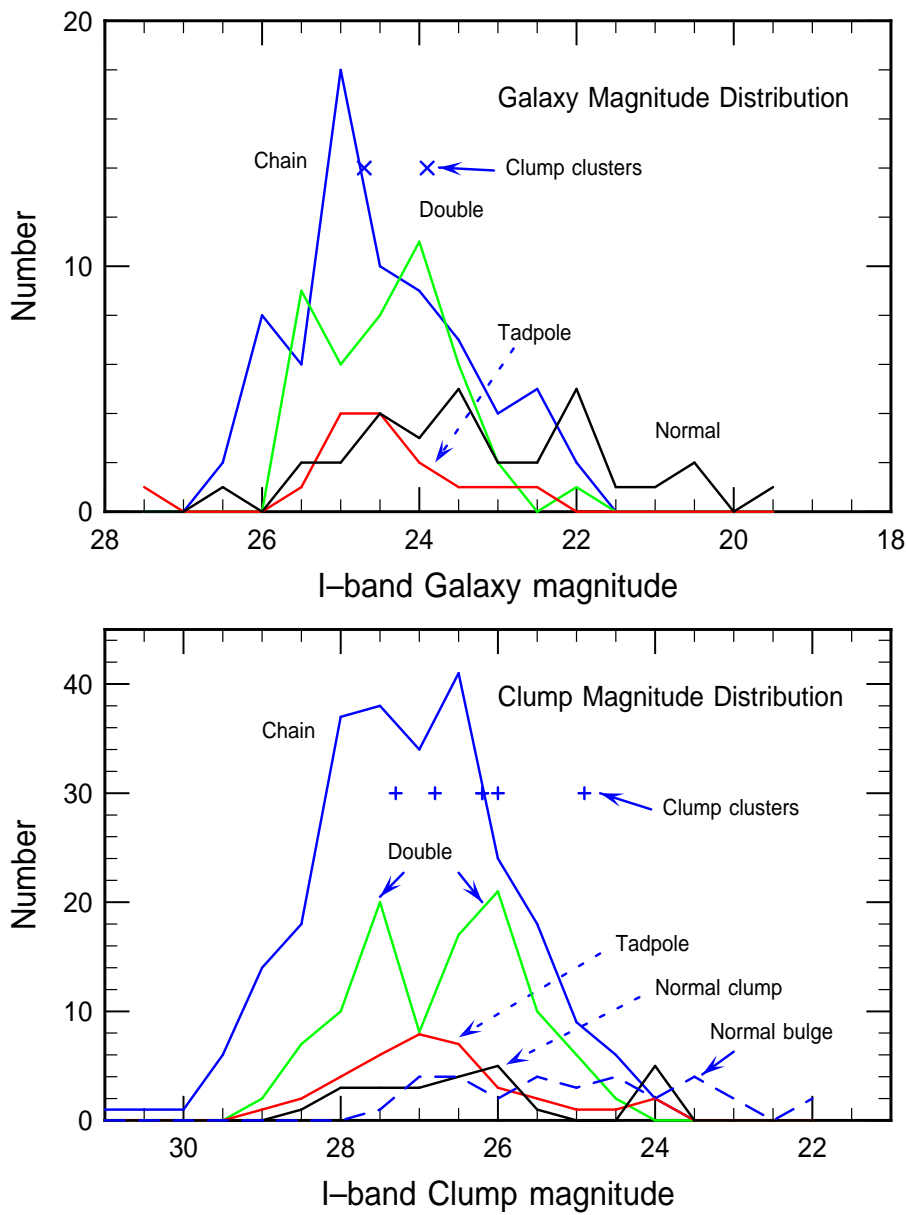} \caption{Distribution of I-band magnitude of
whole galaxies (top) and individual clumps (bottom).  Central
clumps in the normal galaxies are brighter than their non-central
counterparts and brighter than most of the clumps in the linear
galaxies. Whole galaxies are brighter than the clumps by $\sim3$
mag, and most normal edge-on galaxies are brighter than the
chains, doubles, and tadpoles by $\sim2$ mag. The average
brightness of a whole galaxy of the chain, double or tadpole type
is comparable to the average brightness of the bulge in a normal
edge-on galaxy. The plus symbols are the clumps in the clump
clusters shown in fig. 9.}
\end{figure}

\begin{figure}
\plotone{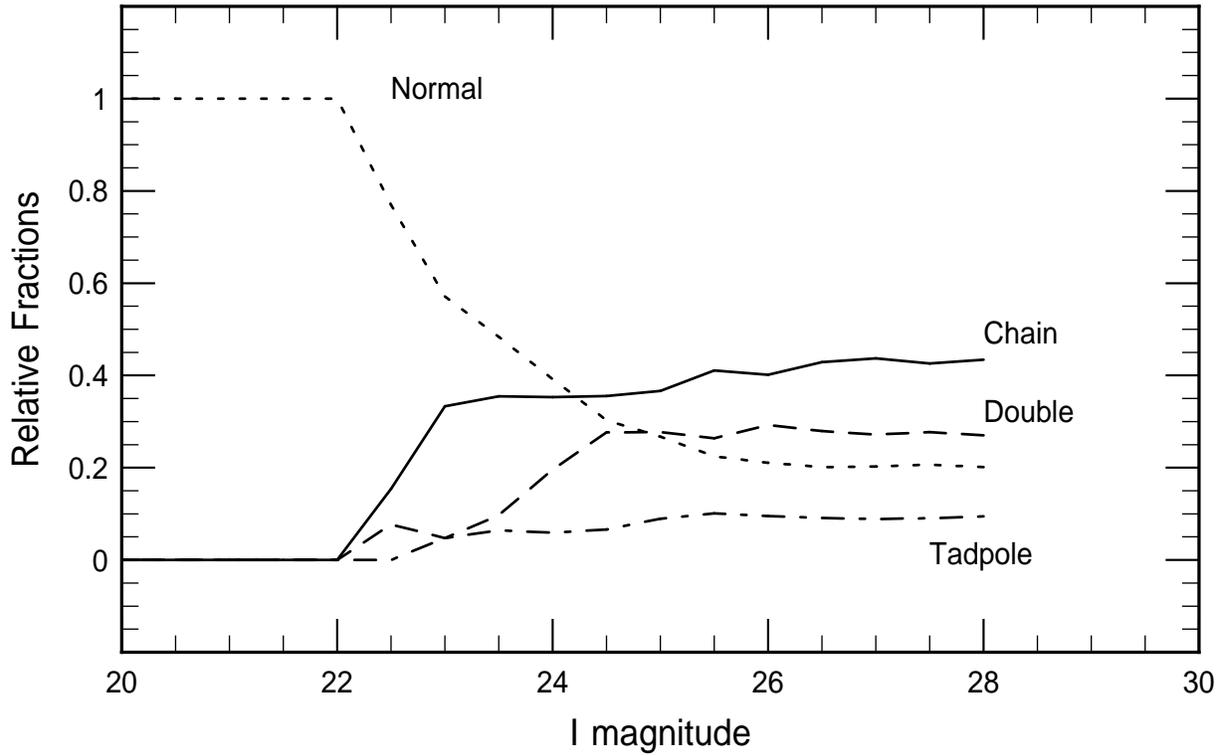} \caption{The fraction of galaxies with a given
morphology is shown as a function of I-band magnitude. The normal
galaxies dominate the bright end; beyond magnitude 24, the chain
galaxies dominate. The fractions are approximately constant beyond
this magnitude.}
\end{figure}

\begin{figure}
\plotone{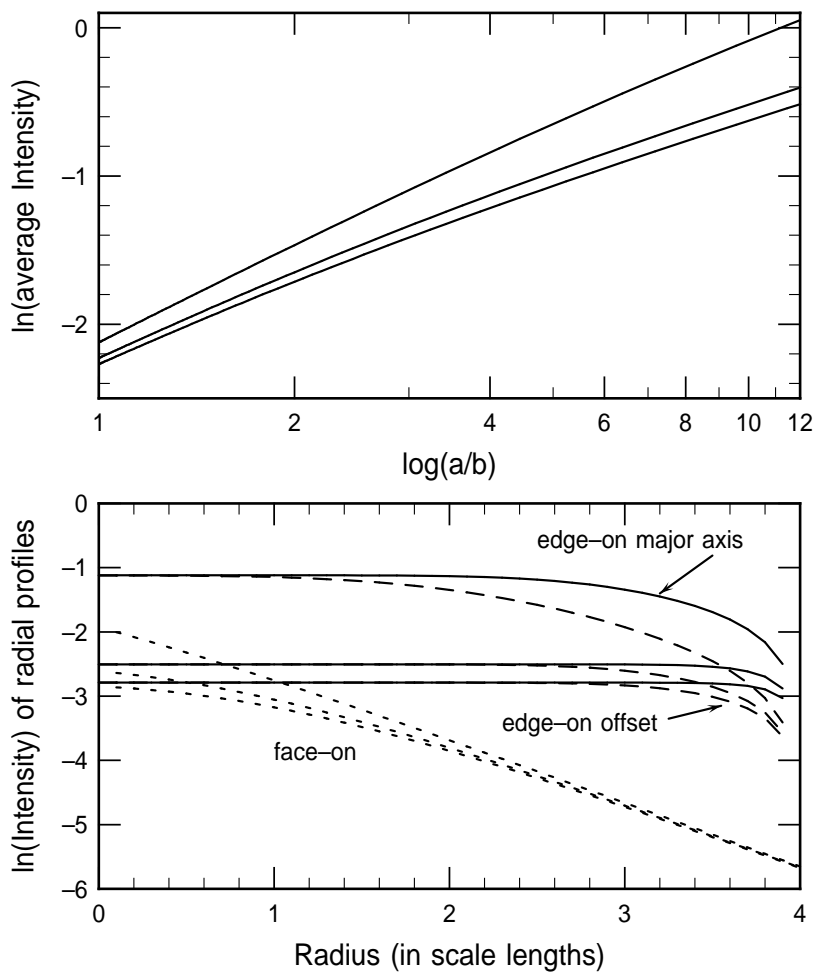} \caption{The bottom panel shows radiative
transfer solutions of the average intensity along the major axis
of a $90^\circ$ inclined dusty disk (solid line) and along a line
displaced from this by one disk scale height (dashed). The dotted
line is the brightness profile for a face-on version of the same
galaxy. The three cases have perpendicular extinctions to the
midplane at the galaxy center equal to 0.33, 1.33, and 1.77
optical depths.  The top panel shows the average surface
brightness of a disk galaxy with an inclination given by the ratio
of axes on the abscissa. Late-type galaxies close to the surface
brightness limit of a survey will be mostly edge-on.}
\end{figure}

\begin{figure}
\caption{Two clump clusters that could be face-on
versions of chain galaxies. The tic marks are separated by 10
pixels ($=0.5$ arcsec). The contours are $1 \sigma$ for the
left-hand clump cluster, as in Fig. 2, and $0.5\sigma$ for the
right-hand cluster. The objects have the same angular size and
clump properties as the chain, double and tadpole galaxies, but
they are 2 dimensional rather than linear. (see 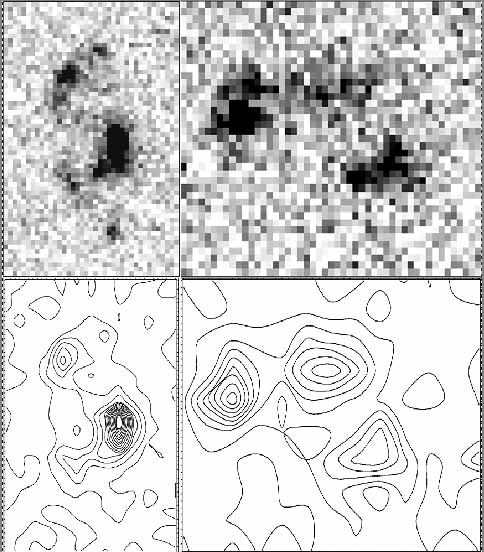)}\end{figure}


\clearpage

\begin{thebibliography}{}

\bibitem[]{467}Abraham, R., van den Bergh, S., Glazebrook, K., Ellis, R.,
Santiago, B., Surma, P., \& Griffiths, R. 1996, ApJS, 107, 1

\bibitem[]{470}Aguerri, J., \& Trujillo, I. 2002, MNRAS, 333, 633

\bibitem[]{472}Blakeslee, J., Anderson, K., Meurer, G., Benitez, N., \& Magee,
D. 2003, in Astronomical Data Analysis Software and Systems XII,
ASP Conference Series, Vol. 295,  H. E. Payne, R. I. Jedrzejewski,
and R. N. Hook, eds., p.257

\bibitem[]{476}Cowie, L.L. Lilly, S.J., Gardner, J., \& McLean, I.S.
1988, ApJ, 332, L29

\bibitem[]{479}Cowie, L., Hu, E., \& Songaila, A. 1995, AJ, 110, 1576

\bibitem[]{481}Cowie, L., Songaila, A., Hu, E. \& Cohen, J.G. 1996, AJ, 112, 839

\bibitem[]{483}Dalcanton, J.J., \& Schectman, S.A. 1996, ApJ, 465, L9

\bibitem[]{485}Hickson, R. 1993, Atlas of compact groups of galaxies.
New York: Gordon and Breach.

\bibitem[]{486} Holmberg, E. 1958, Medd. Lunds. Astr. Obs., Ser. 2,
136

\bibitem[]{487}Koo, D., Voot, N., Phillips, A., Guzman, R., Wu, K., Faber, S.,
Gronwall, C., Forbes, D., Illingworth, G., Groth, E., Davis, M.,
Kron, R., \& Szalay, A. 1996, ApJ, 469, 535

\bibitem[]{491}Kravtsov, A.V. \& Gnedin, O.Y. 2003, astroph/0305199

\bibitem[]{493}Lilly, S.J., Cowie, L.L., \& Gardner, J.P. 1991, ApJ, 369, 79

\bibitem[]{495}McGaugh, S., \& Bothun, G. 1994, AJ, 107, 530

\bibitem[]{} Mendes de Oliveira, C. \& Hickson, P. 1994, ApJ, 427, 684

\bibitem[]{497}Metcalfe, N., Shanks, T., Fong, R., \& Roche, N. 1995, MNRAS,
273, 257

\bibitem[]{500}O'Neil, J., Bothun, G.D., \& Impey, C.D. 2000, ApJS, 128, 99

\bibitem[]{502}Pohlen, M., Balcells, M., Lutticke, R., \& Dettmar, R.-J. 2003,
A\&A, 409, 485

\bibitem[]{505}Reshitnikov, V., Dettmar, R.-J., \& Combes, F. 2003, A\&A, 399,
879

\bibitem[]{} Solanes, J.M., Giovanelli, R., \& Haynes, M.P. 1996,
ApJ, 461, 609

\bibitem[]{508}Steidel, C.C., Giavalisco, M., Dickinson, M., \& Adelberger,
K.L. 1996, AJ, 112, 3525

\bibitem[]{511}Taniguchi, Y., \& Shioya, Y. 2001, ApJ, 547, 146

\bibitem[]{513}Tran, H. et al., 2003, ApJ, 585, 750

\bibitem[]{515}van den Bergh, S., Abraham, R.G., Ellis, R.S., Tanvir, N.R.,
Santiago, B.X., \& Glazebrook, K.G. 1996, AJ 112, 359

\bibitem[]{518}van den Bergh, S., Cohen, J.G., Hogg, D.H., \&
Blandford, R. 2000, AJ, 120, 2190

\bibitem[]{521}Volonteri, M., Saracco, P., \& Chincarini, G. 2000, A\&AS, 145,
111

\bibitem[]{524}Vorontsov-Velyaminov, B.A. 1977, A\&AS, 28, 1

\bibitem[]{526} Westera, P., Samland, M., Buser, R., \& Gerhard, O. 2002, A\&A
389, 761

\bibitem[]{529}Williams, R., et al. 1996, AJ, 112, 1335

\end{thebibliography}
\end{document}